\documentclass[letterpaper, 10 pt, conference]{ieeeconf}  

\IEEEoverridecommandlockouts                              

\usepackage[utf8]{inputenc}

\usepackage{graphicx,comment,cite}
\usepackage{amsmath,amssymb,amsthm,soul} 
\usepackage{graphicx,cite}
\usepackage{array}

\newtheorem{theorem}{Theorem}
\newtheorem{lemma}{Lemma}

\usepackage{xcolor}
\usepackage{framed}
\usepackage{lipsum}
\colorlet{shadecolor}{yellow!80}

\title{\LARGE \bf Straggling for Covert Message Passing on Complete Graphs}

\author{Pei Peng$^{1}$, Nikolas Melissaris$^{1}$, Emina Soljanin$^{1}$, Bill Lee, Anton Maliev, and Huafeng Fan
\thanks{$^{1}$The authors are with the Department of Electrical and Computer Engineering,
        Rutgers University, the State University of New Jersey, USA
        {\tt\small \{pei.peng,nikolas.melissaris, emina.soljanin\}@rutgers.edu}}%
}
\begin{document}

\maketitle

\begin{abstract}
We introduce a model for mobile, multi-agent information transfer that increases the communication covertness through a protocol which also increases the information transfer delay. Covertness is achieved in the presence of a \textit{warden} who has the ability to patrol the communication channels. Furthermore we show how two forms of redundancy can be used as an effective tool to control the tradeoff between the covertness and the delay. 
\end{abstract}
\section{Introduction}
\label{Sec:intro}
In numerous circumstances, the very act of communicating needs to be hidden. In war time, not only the content of messages, but also the volume of communication to or from suspected parties can alert the adversary. In everyday life, revealing the identity of communicating parties, not only the exchanged information, affects the increasingly important anonymity and privacy. An information theoretic approach to achieving covertness, adopted in several recent papers (see e.g. \cite{HidingInformationInNoise:BashGT15, CovertComm:Bloch16, DeniableComm:KadheJB14} and references therein), roughly speaking, relies on camouflaging messages as noise. 
We here propose covert message passing in a wireless mobile network environment that is complementary to and can be used in conjunction with the previously proposed methods.

The last decade has seen a wide variety of novel networked systems, with a growing trend towards distributed and multiuser applications. Future 5G systems are supposed to host hundred times more devices than current 4G networks, and one can potentially harness the resources expected to be brought in by smart (everyday or battlefield) devices in the future Internet of (Battlefield) Things (Io$\beta$T) environments. 

The precise mathematical description of our problem will be presented in the following section; we next give two high level examples.
Consider a scenario where an agent Alice has a message she wants to covertly send to her partner Bob who resides in the same (possibly occupied) city. We represent city streets as a graph. Information gathering and dissemination on graphs is a very interesting problem that arises naturally and has recently seen active research in many different contexts. Examples include: border control using UAVs \cite{girard2004border}, measuring traffic, reporting road conditions and helping with emergency response using UAVs \cite{puri2005survey}, monitoring the ocean \cite{paley2008cooperative}, measuring air pollution \cite{villa2016development}, and more recently for timely exchange of information updates \cite{DBLP:conf/infocom/TripathiTM19}.

The area over which Alice and Bob plan to communicate hosts a multitude of IoT objects capable of storing, sending and receiving data.
Alice can use some of these smart objects to store her data, and Bob can subsequently retrieve the stored data. The agents have one or both of the following concerns: 1) they do not want to be detected  communicating with any of the relay objects and 2) the system should be trustworthy even if some of the objects e.g., have lost power, are unwilling to relay messages, or are otherwise compromised in an adverse environment. 
Because of that, Alice decides to split her data into small chunks which she can  inconspicuously pass, one at the time, to relays (helpers) that appear in her proximity as she randomly moves through the area. Bob, who also randomly moves through the area, can then retrieve the stored data chunks. We assume that there is a warden Willie patrolling the city. Because the IoT objects are distributed in a wide area, Willie can only periodically check if any of these objects is transmitting or receiving data. 

Having to distribute and collect many chunks, as well as 
the unpredictability of mobility and availability of helpers can cause large delays in such mobile information transfer. 
To increase persistence of information in the erratic environment, the agent may decide to make the data chunks redundant by using erasure correcting codes, which requires that more data chunks be distributed and collected. One would expect that that would further increase  the delay in mobile information transfer. However, that is not necessarily the case, and we will see that coding and some other forms of redundancy can in fact be used to reduce the delay, as previously shown to be the case in data download \cite{joshi2012coding}.

There is another analogy for our communications scenario: Alice and Bob are at a party and Alice wants to send a message to Bob without actually talking to him. Willie, who is also at the party, wants to detect whether Alice and Bob are communicating. Willie could easily detect if Alice talks directly with Bob, so Alice decides to divide the message into several pieces and whispers each piece to a different mutual friend she has with Bob. Bob can then walk around in the party and collect these message pieces from the friends to retrieve the message. Since Willie is unable to keep track of all the participants in the party, Alice may be able send the message successfully to Bob without being detected. 

If the size of the message that Alice wants to send to Bob is fixed,
then the delay and the covertness of the message transfer will depend on the number of chunks that the message is split into and the amount of redundancy (code rate) that is introduced. We here derive the expressions that describe this dependence under certain communication model.
The paper is organized as follows: In Sec.~\ref{Sec:sys}, we present the system model and point out multiple tradoffs that exist between the time to disseminate data, time to collect data, and the probability of covertness. In Sec.~\ref{Sec:analysis}, we derive expressions for the covertness probability and dissemination/collection delays. 
In Sec.~\ref{Sec:perform}, we present some numerical results and analyze the maximum covertness probability and minimum delay by simulation. 

\section{System Model}
\label{Sec:sys}

\subsection{Communications Protocol \& the Mobility Model}
There are four (types of) participants in our communication model: a source (Alice), $r$ relays, a collector (Bob), and a warden (Willie). Alice and Bob walk randomly on a complete graph of $s$ vertices. The $r$ relays are placed on $r$ different vertices of this graph. 
Alice has a message of length $m$ which she divides into $k\le m$ chunks of length $\ell=m/k$, and then encodes the data chunks into $n\le r$ coded chunks (of length still $\ell$). She transfers the $n$ coded chunks to the first $n$ relays she encounters as she randomly walks through the graph. The collector Bob retrieves the message by collecting $k$ chunks from the first $k$ relays with data chunks he encounters as he randomly walks through the graph. 
Note that we assumed without loss of generality that the encoding is such that any $k$ out of $n$ chunks are sufficient for message recovery. We assume a complete graph for our communication graph because it simplifies the analysis and drives our point home without getting into messy calculations. Other communication graphs can be used as well in future research, especially large regular graphs or graphs with a few high degree hubs.

Notice how the tradeoff between covertness and delay becomes apparent here. On one hand, if Alice sets $n=1$, meaning that she delivers the entire message in one chunk, then the probability of detection is small. This happens because the chance that Willie ``sees'' her is inversely proportional to the number of nodes. On the other hand, delay is increased because Bob will have to visit many nodes until he meets the relay that holds Alice's message.

At the other side of the spectrum if Alice splits the message in many chunks, then it will take Bob less steps to retrieve it (since the probability of meeting a relay that holds a message is high) but the probability that they are caught increases significantly because of the longer time that they have to spend walking on the graph.

\subsection{Chunk Transmission}
We assume that the transmission time of a chunk (between Alice and a relay or Bob and a relay)  
consists of two additive components. The first is proportional to the chunk length $\ell$, and the other is an independent random variable that accounts for various disturbances (noise) in the system. We will assume that this random variable is exponential with parameter $\lambda$, and thus the transmission time is shifted exponential with the shift value $\ell$ and the tail 
parameter $\lambda$.

\subsection{Warden Models}
We now introduce our very simple warden model. We base covertness on the assumption that the warden Willie is able to monitor part of the vertices for some given time. A very easy and informal way to visualize this is to imagine that the warden is stationed somewhere ``in the middle'' of the graph\footnote{This is a very informal statement. There is no need for an exact ``middle'', we just need the warden to be at a place where he can observe different parts of the network at different times.}, on top of a lighthouse. This way, he can only check the parts where the lighthouse sheds its light and can't see what is happening behind him. We assume that no transmission covertness scheme is implemented, and thus if the warden happens to check a vertex of the graph while the chunk transmission is taking place, he will detect it with probability 1. We further assume that when the transmission starts, the warden's (monitoring) time follows a uniform distribution $U(0,W)$.  The information transfer stays covert iff the transmission of each chunk stays undetected. There are multiple warden models that can be considered in future research. In this work, we observe two simple ones in order to point out the tradeoff between covertness and delay that our protocol manages to balance. The main idea separating the models is whether Willie can learn something about where meaningful transmission takes place just by observing how Alice and Bob walk along the graph. More details are made explicit in sections \ref{subsec:Model1} and \ref{subsec:Model2}.
 

\section{Performance Metrics}
Our performance metrics of interest are \emph{covertness probability}, the \emph{dissemination time} of the $n$ coded chunks by Alice, the \emph{collection time} of $k$ coded chunks by Bob, and the total data transfer time (dissemination plus collection). We will see, in the following sections, that each performance metric is optimized by a different set of system parameter values. 
\subsection{Dissemination/Collection Time}

We can consider the dissemination/collection time as a coupon collector's problem: there are $n$ stores at city squares, each one selling a different coupon, and there is a direct road between any two squares. A coupon collector walks randomly in the city and want to buy $j$ different coupons. If $j=1$, it is obvious the collector needs to visit only $1$ square get a coupon; If $j=2$, he needs to visit on average $1+\frac{n}{n-1}$ squares; If $j=3$, the average visited squares is $1+\frac{n}{n-1}+\frac{n}{n-2}$. Following this pattern, we know that when $j=n$, the collector needs to visit on average $nH_n$ squares, where $H_{n}=\sum_{i=1}^{n}1/i$ is the $n$-th harmonic number. During each visiting, if he spends on average $T_{ave}$ in the store, the total average time is $nT_{ave}H_n$. The dissemination time and collection time can be calculated in a similar way.

\subsection{Covertness Probability}
The covertness probability is defined as the probability that Alice transmits a message to Bob without being detected by Willie. For example, assuming the message has $2$ data chunks, then Alice needs to transmit $2$ times to relays and Bob also needs $2$ times to collect the chunks. If during each time, Willie will detect the transmission with a probability $P_d$, then the covertness probability is $P_c=(1-P_d)^4$. Notice that when Willie detects the transmission, it doesn't means he will get the content of the message. The message may be camouflaged as noise to avoid detection. But this is another covert communication problem, and won't be studied in this paper.

\section{Theoretical Performance Analysis}
\label{Sec:analysis}

\subsection*{System Parameters}
\begin{center}
    \begin{tabular}{rcl}
       $s$ & -  &  number of graph vertices \\
       $r$ & -  & number of relays\\
       $m$  & -  & length of the message in bits\\
    $k$  & -  & number of message (data) chunks \\
    $n$  & -  & number of encoded chunks \\
    $\ell$  & -  & length of the chunk in bits
    \end{tabular}
\end{center}

\subsection{Covertness Probability}
\label{sec:covert}
The detection probability of each single chunk transmission is given by the following theorem:

\begin{theorem}
\label{Th:covert2}
If the transmission time between the source (collector) and a relay follows a shifted exponential distribution $\lambda e^{-\lambda (t-\ell)}$ and the Willie's monitoring arrival time at the vertex has a uniform PDF $U(0,W)$, then the probability that Willie arrives during the transmission (i.e., detects the transmission) is 
\[
    P_d=\begin{cases}
             \frac{1}{\lambda W}+\frac{m}{kW}-\frac{e^{-\lambda (W-m/k)}}{\lambda W} & \text{for }  W \ge \ell \\
               \hfil 1  & \text{for }  W < \ell
\end{cases}
\]
\end{theorem}
\noindent{\it Proof:}\space
If $W<\ell=m/k$, then Willie will definitely arrive before the transmission is complete, and thus the detection probability is $P_d=1$. 
\\[1ex]
If $W> \ell$, we calculate the detection probability as follows:
\begin{small}
    \begin{align*}
        P_d=P(t_{tr}\ge t_{ar})&=\int^{\infty}_{0}\int^{\infty}_{t_{ar}}f_{tr}(t_{tr})f_{ar}(t_{ar}) \ dt_{tr} \ dt_{ar}\\
        &=\int^{\infty}_{\frac{m}{k}}\int^{\infty}_{t_{ar}}f_{tr}(t_{tr}) \ dt_{tr} \ f_{ar}(t_{ar}) \ dt_{ar}\\&+\int^{\frac{m}{k}}_{0}\int^{\infty}_{\frac{m}{k}}f_{tr}(t_{tr}) \ dt_{tr} \ f_{ar}(t_{ar}) \ dt_{ar}\\
        &=\int^{\infty}_{\frac{m}{k}} e^{-\lambda (t-m/k)}f_{ar}(t_{ar}) \ dt_{ar}+\frac{m}{kW}\\
        &=\int^{W}_{\frac{m}{k}} \frac{1}{W} e^{-\lambda (t-m/k)} \ dt_{ar}+\frac{m}{kW}\\
        &=\frac{1}{\lambda W}+\frac{m}{kW}-\frac{e^{-\lambda (W-m/k)}}{\lambda W}
    \end{align*}
\end{small}

The entire message transmission will be undetected if each chunk transmission is undetected. Recall that Alice needs to disseminate the $n$ chunks and Bob needs to collect $k$ chunks. Therefore the total number of chunk transmissions is $n+k$, and thus the overall covertness probability is 
\begin{equation}
     P_c= (1-P_d)^{n+k}
 \label{Le:covert}   
\end{equation}

From Theorem \ref{Th:covert2} and  equation (\ref{Le:covert}), we can see that when $n$ increases, the covertness probability decreases, which means that coding redundancy hurts the covertness. However, it is less clear how changing the number of chunks $k$ affects the covertness probability, as we discuss below.

\section{Communication Delay}

The message transfer from Alice to Bob consists three stages: 1. the meeting time steps: the source and collector need to meet enough relays; 2. The discovering time: the source and collector need to find out if there is a relay on the same vertex; 3. The transmission time: the dissemination/collection time between the source/collector and the relay. As we know, the transmission time relates to the size of the data chunks. It's reasonable to assume that the larger data chunk needs to spend a longer transmission time.

To calculate the communication delay in covert communication, we also need to consider the Willie. Based on different Willies's detection patterns, we provide two communication delay models. 

For the first model, if Willie can't observe the movements of Alice and Bob, then he can not simply detect the transmission by comparing the time Alice/Bob spends on each vertex. Therefore, we assume that during the source's random walk we can have two cases: Either the source does not meet a relay and the discovering time is $1$ or the source meets a relay on a vertex and we have a transmission time $1+t_{tr}$. As described in Sec.~\ref{sec:covert}, we model the transmission time $t_{tr}$ as a shifted exponential random variable with the tail parameter $\lambda$ and the shift parameter $m/k$.  

For the second model, if Willie can observe the movements of Alice and Bob, he may learn which nodes they are stopping for longer times at, and conclude that they hold valuable information. To prevent this, one idea is to spend an equal length of time at each visited node, and even transmitting an empty signal with no valuable information on non-relay vertices. This way, even if Willie can track Alice's and Bob's movements, he does not gain any additional information about their communication. Now, every node visited by Alice or Bob takes $1+m/k$ units of time.

\subsection{Model 1}
\label{subsec:Model1}

\subsubsection{Dissemination Time}

In the dissemination stage, the source Alice needs to disseminate $n$ chunks to $r$ relays. When Alice randomly walks on a complete graph with $s$ vertices, the probability that she meets a relay is $r/s$. After Alice deposits the first data chunk in one of the $r$ relays, the second chunk can only be stored in one of the remaining $r-1$ relays. The probability of meeting an occupied relay decreases as the  number of occupied relays grows. Therefore, in order to get the total dissemination time $T_{\tt dis}$, we need to find dissemination time $T_i$  where $i=\{1,2,\dots,n\}$ for each 
encoded chunk of the message. 

\begin{lemma}
\label{Th:dis-time}
The dissemination time $T_i$ to transmit the $i^{th}$ data chunk to any one of $r-i+1$ relays is
\begin{equation}
    T_i=t_{tr}+1+a \quad \text{ for a probability } (1-p_{r-i+1})^{a}p_{r-i+1} 
\end{equation}
Where $a \in \{0,1,2,...\}$ is the number of steps the source spent to meet a relay, $p_{r-i+1}=\frac{r-i+1}{s}$.

And then the expectation of $T_i$ is
\begin{equation}
   \mathbb{E}\left[T_i\right]= \frac{1}{\lambda}+\frac{m}{k}+\frac{1}{p_{r-i+1}}
\end{equation}
\end{lemma}

\noindent{\it Proof:}\space
Since for time $T_i$, the value of $p_{r-i+1}$ is a constant. Let's assume $p=p_{r-i+1}$.
\begin{small}
    \begin{align*}
        \mathbb{E}\left[T_i\right]&=\sum_{a=0}^{\infty}\mathbb{E}\left[t_{tr}+1+a\right](1-p)^{a}p\\
        &=p(\mathbb{E}\left[t_{tr}\right]+1)\sum_{a=0}^{\infty}(1-p)^{a}+p(1-p)\sum_{a=1}^{\infty}a(1-p)^{a-1}\\
        &=\mathbb{E}\left[t_{tr}\right]+\frac{1}{p}
        =\frac{1}{\lambda}+\frac{m}{k}+\frac{1}{p_{r-i+1}}\qquad\qed
    \end{align*} 
\end{small}

\begin{lemma}
\label{Lm:dis-expect}
The total dissemination time $T_{\tt dis}$ is that the source transmits all $n$ chunks to any $n$ out of $r$ relays. Then the expectation of time $T_{\tt dis}$ is 
\begin{equation}
    \mathbb{E}\left[T_{\tt dis}\right]=\frac{n}{\lambda}+\frac{nm}{k}+s(H_r-H_{r-n})
\end{equation}
\end{lemma}
\noindent{\it Proof:}
\begin{small}
    \begin{align*}
        \mathbb{E}\left[T_{\tt dis}\right]&=\sum_{i=1}^{n}T_i
        =n\mathbb{E}\left[t_{tr}\right]+\sum_{i=1}^{n}\frac{1}{p_{r-i+1}}\\
        &=n\mathbb{E}\left[t_{tr}\right]+\sum_{i=1}^{n}\frac{s}{r-i+1}\\
        &=\frac{n}{\lambda}+\frac{nm}{k}+s(H_r-H_{r-n})\hfill \qed
    \end{align*}
\end{small}

It's easy to see that $n$ reaches optimal at $n=k$, which means the less redundancy, the lower dissemination delay.

\subsubsection{Collection Time}

During the collection step, the collector needs to collect any $k$ out of $n$ data chunks to recover the message. Since the transmission time between the collector and relay follows the same distribution as the time in dissemination step. Then as in Lemma \ref{Lm:dis-expect}, we can find the total expectation of collection time $T_{\tt col}$ in Lemma \ref{Lm:col-expect}.

\begin{lemma}
\label{Lm:col-expect}
The total collection time $T_{\tt col}$ is the time spent by the collector to retrieve any $k$ chunks from $n$ relays. Then the expectation of time $T_{\tt col}$ is 
\[
    \mathbb{E}\left[T_{\tt col}\right]=\frac{k}{\lambda}+m+s(H_n-H_{n-k})
\]
\end{lemma}

From Lemma \ref{Lm:col-expect}, $\mathbb{E}\left[T_{\tt col}\right]$ is decreasing when $n$ is increasing.

\subsubsection{Joint time}
After getting the dissemination time and collection time, the joint time is just the sum of these two times.
\begin{theorem}
\label{Lm:joi-expect}
Now we can easily calculate the expectation of the joint time:
\[
    \mathbb{E}\left[T_{tot}\right]=\frac{n+k}{\lambda}+\left(\frac{n}{k}+1\right)m+s(H_r+H_n-H_{r-n}-H_{n-k})
\]
\end{theorem}

From the conclusions in dissemination and collection steps, we can see that there must be an optimal $n$ which minimize the $\mathbb{E}\left[T_{tot}\right]$. To calculate the optimal $n$ is complicated, so we will show result by simulation in next section.

\subsection{Model 2}
\label{subsec:Model2}

\subsubsection{Dissemination time}
\begin{lemma}
\label{Th:dis-time-2}
The total dissemination time $T_{\tt dis}$ is the time needed to transmit $n$ code blocks to any $n$ out of $r$ relays. Then the expected value of $T_{\tt dis}$ is:
\begin{equation}
    \mathbb{E}[T_{\tt dis}]=\left(\frac{1}{\lambda}+\frac{m}{k}+1\right)s\left(H_r-H_{r-n}\right)
\end{equation}
\end{lemma}

\noindent{\it Proof:}\space
We follow the same reasoning as for the original model.
The expected number of nodes visited can be computed by modeling the relay traversal as coupon-collecting: vising $n$ out of a given $r$ nodes out of $s$ total nodes. The expected number of visits needed is $s(H_r-H_{r-n})$. At every node visited, 1 unit of time is needed to check if the node is a relay, and the transmission takes an expected time of $1/\lambda+m/k$. 
\medskip

\subsubsection{Collection time}

\begin{lemma}
\label{Th:col-time-2}
The total collection time $T_{\tt col}$ is the time needed to collect $k$ code blocks from any $k$ out of $n$ relays. Then the expected value of $T_{\tt col}$ is:
\begin{equation}
    \mathbb{E}[T_{\tt col}]=\left(\frac{1}{\lambda}+\frac{m}{k}+1\right)s\left(H_n-H_{n-k}\right)
\end{equation}
\end{lemma}

This result is obtained in the same manner as the dissemination time, except that Bob needs to visit any $k$ out of $n$ with data on a graph with $s$ nodes. 

\medskip

\subsubsection{Joint Time}

\begin{theorem}
\label{Col:joint-time-2}
The expected value of the total time needed for dissemination and collection is
\[
    \mathbb{E}[T_{tol}]=\left(\frac{1}{\lambda}+\frac{m}{k}+1\right)s\left(H_r+H_n-H_{r-n}-H_{n-k}\right)
\]
\end{theorem}

For this model, we next derive an expression for the optimal value of $n$ that minimizes the total transmission time, and omit the simulations.

For given $k$, $\lambda$, $m$, $s$ and $r$, we denote $A_n=H_n-H_{r-n}-H_{n-k}$. Then we have $A_n-A_{n+1}=-\frac{1}{n+1}-\frac{1}{r-n}+\frac{1}{n+1-k}$.\\
If $A_n-A_{n+1}<0$, we have
\begin{small}
\begin{align*}
 &-\frac{1}{n+1}-\frac{1}{r-n}+\frac{1}{n+1-k}<0\\
 &\Leftrightarrow n^2+2n+1-rk-k>0\\
 &\Leftrightarrow n>\sqrt{rk+k}-1
\end{align*}
\end{small}
Similarly, if $A_n-A_{n+1}>0$, then 
$
n<\sqrt{rk+k}-1.
$
Thus we can optimize the total delay by producing $\left \lceil{\sqrt{rk+k}-1} \right \rceil$ or $\left \lfloor{\sqrt{rk+k}-1} \right \rfloor$ code blocks.

\section{Numerical and Simulation Analysis}
\label{Sec:perform}
\subsection{Numerical Covertness Probability Analysis}
Since it is easy to see that the covertness probability increases with the redundancy $n$, we only discuss how the covertness probability changes with $k$. We consider an example where the distribution of transmission time is shifted exponential with the tail parameter $1$ and shift parameter $10/k$. The Willie's arrival time follows uniform distribution $U(0,50)$. 

\begin{figure}[t]
    \centering
    \includegraphics[scale=0.199]{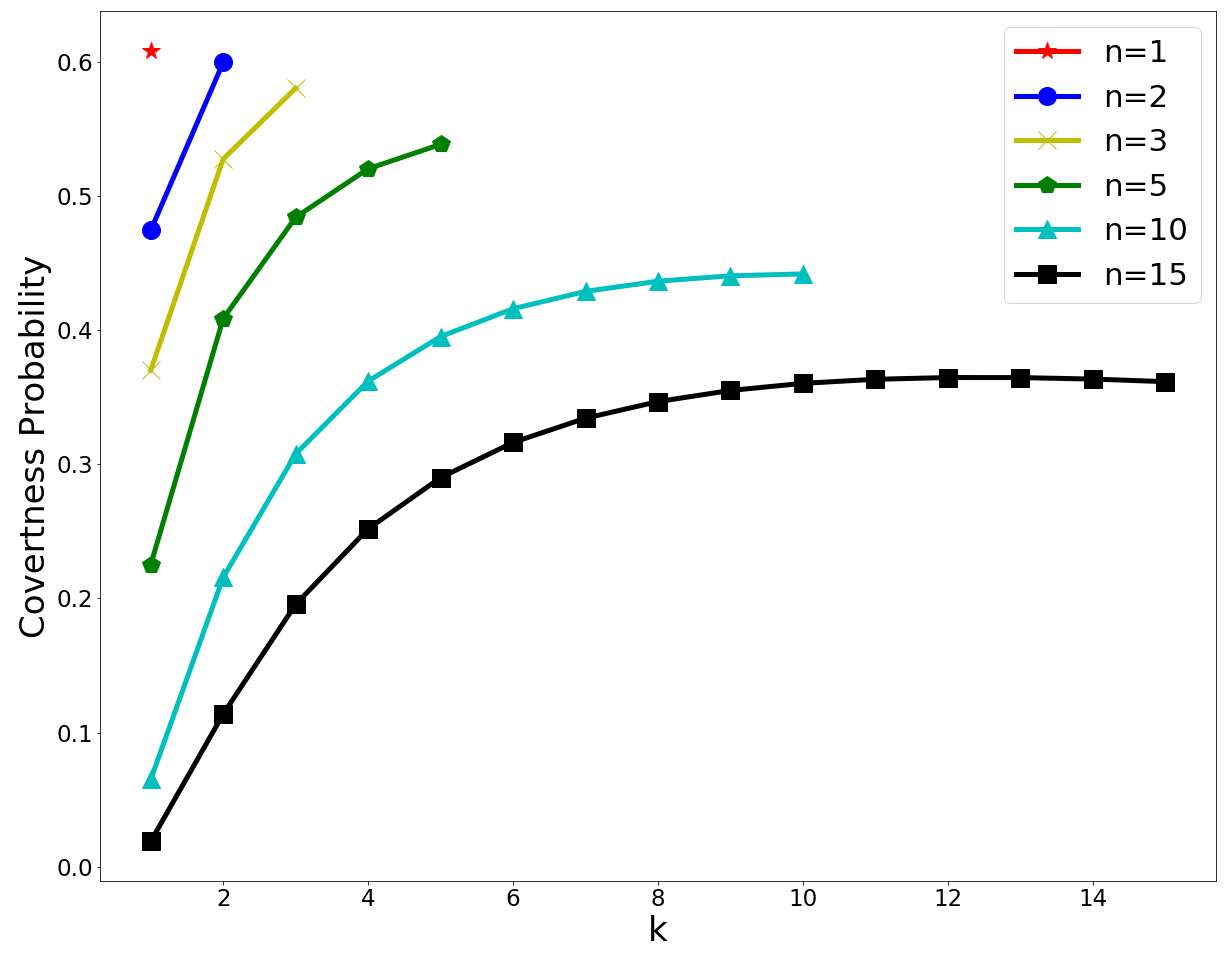}
    \caption{The covertness probability vs.\ $k$ for 5 different values of $n$. The covertness probability increases rapidly with $k$ when $k$ is small. Then the benefits of increasing $k$ get smaller and even negative.}
    \label{fig:covertness}
\end{figure}

Figure~\ref{fig:covertness} plots the covertness probability vs.\ $k$ for 5 different values of $n$ for this example.
We see from the figure that, when $n$ is small (e.g. $n=3$), the covertness probability rapidly increases with $k$, and it is better to select $k$ as large as possible. As $n$ is increasing (cf.\ $n=10$), the covertness probability essentially stops increasing with $k$ after some point (around $k=6$). At that point, there is no need to increase $k$. When $n$ is large (e.g. $n=15$), the covertness probability decreases with $k$ after some point ($k=12$ in the figure). Thus there is an optimal $k$. Also we can see when $n=1$, which means no redundancy is introduced, the covertness probability reaches the maximum. However when $n=2$ or $n=3$ Since the redundancy may provide may reduce the communication delay between the source and collector, it is worth to select a lower covertness probability scenario.

We also notice that the covertness probability in the figure is very low. It is because of the simulation parameters' values we selected. If we change the values, e.g. the Willies's arrival time follows uniform distribution $U(0,100)$, the covertness probability will increase significantly. Since our covert communication model is a new model, we don't know the exact parameters' values in practice. Besides, in this simulation, we want to study how covertness probability changes with the number of message chunks. Therefore the changing of covertness probability is more important than its values.

\subsection{Minimum Delay Analysis by Simulation}

We simulated our message passing protocol on a complete graph with 50 vertices. Again, the transmission time follows shifted exponential distribution with scale parameter $1$ and shift parameter $10/k$. We are interested in seeing how the redundancy parameter $n$ affects the average message transfer time for 1) different values of $r$ and 2) different values of $k$. We obtained the results by simulation and also computed the corresponding theoretical results by using Theorem~\ref{Lm:joi-expect}. 

\begin{figure}[t]
    \centering
    \includegraphics[scale=0.2]{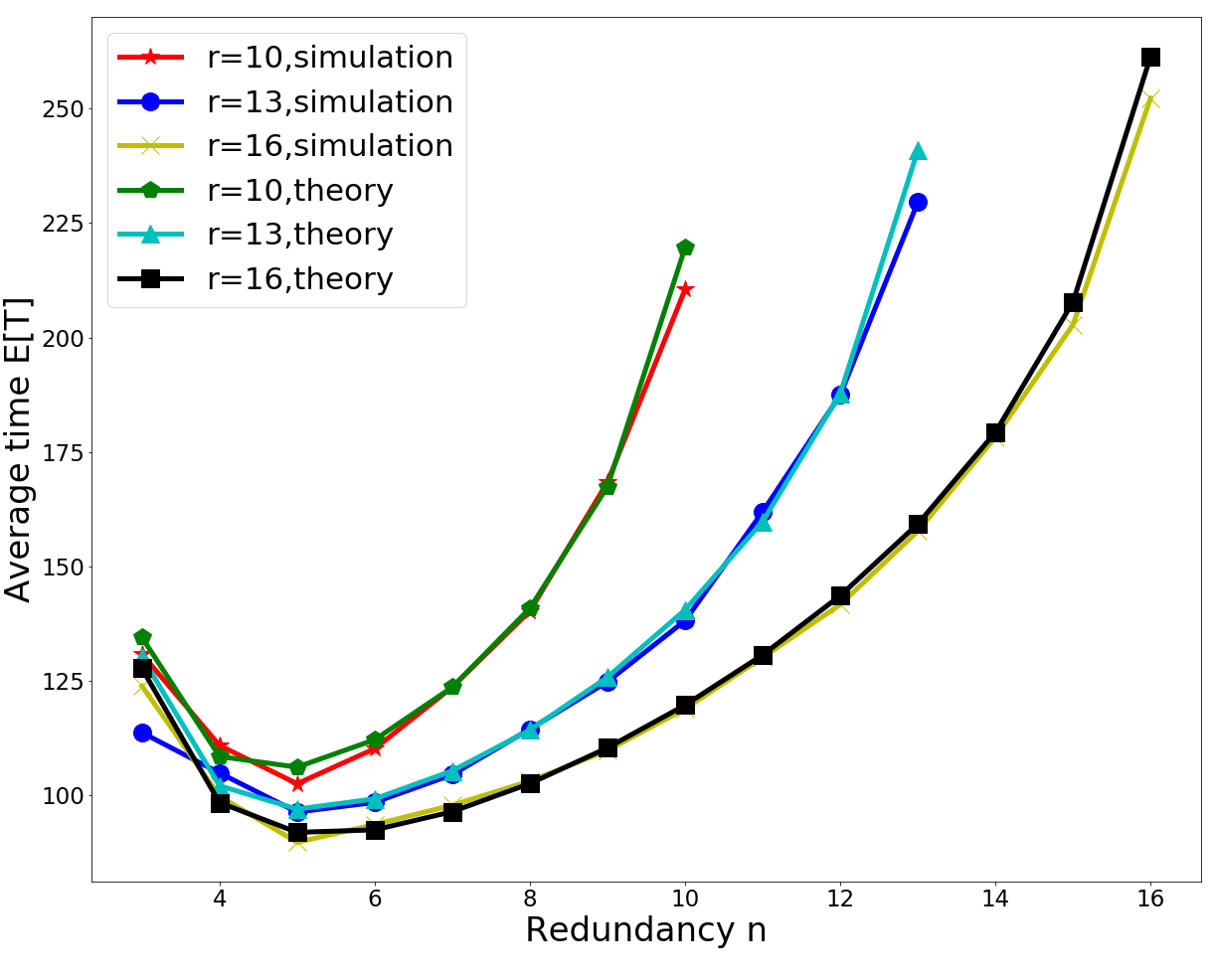}
    \caption{The average time changes with redundancy $n$ under different $r$. The number of message chunks $k=3$. Both simulation and theoretical results are provided.}
    \label{fig:mredundancy}
\end{figure}

{\it 1) Delay vs.\ $n$ for Different Values of $r$:}
Figure \ref{fig:mredundancy} shows both the simulated and the theoretical results, which closely match each other.
From the figure, we can draw several conclusions. First, when we consider each curve in the figure, we can see that introducing proper redundancy can significantly reduce the average time, but too much redundancy can hurt the performance. The optimal redundancy is reached at $n=5$. Second, when we compare all the simulation results or theoretical results, increasing the number of relays $r$ can reduce the average time. This conclusion is not hard to understand. As we know the optimal redundancy is the trade-off between the dissemination time and collection time. When we have more relays, the dissemination time will get reduced and the collection time will not be affected. Then we may increase dissemination time to reduce the collection time. This can be easily derived from the equation in Theorem \ref{Lm:joi-expect}. 

\begin{figure}[t]
    \centering
    \includegraphics[scale=0.195]{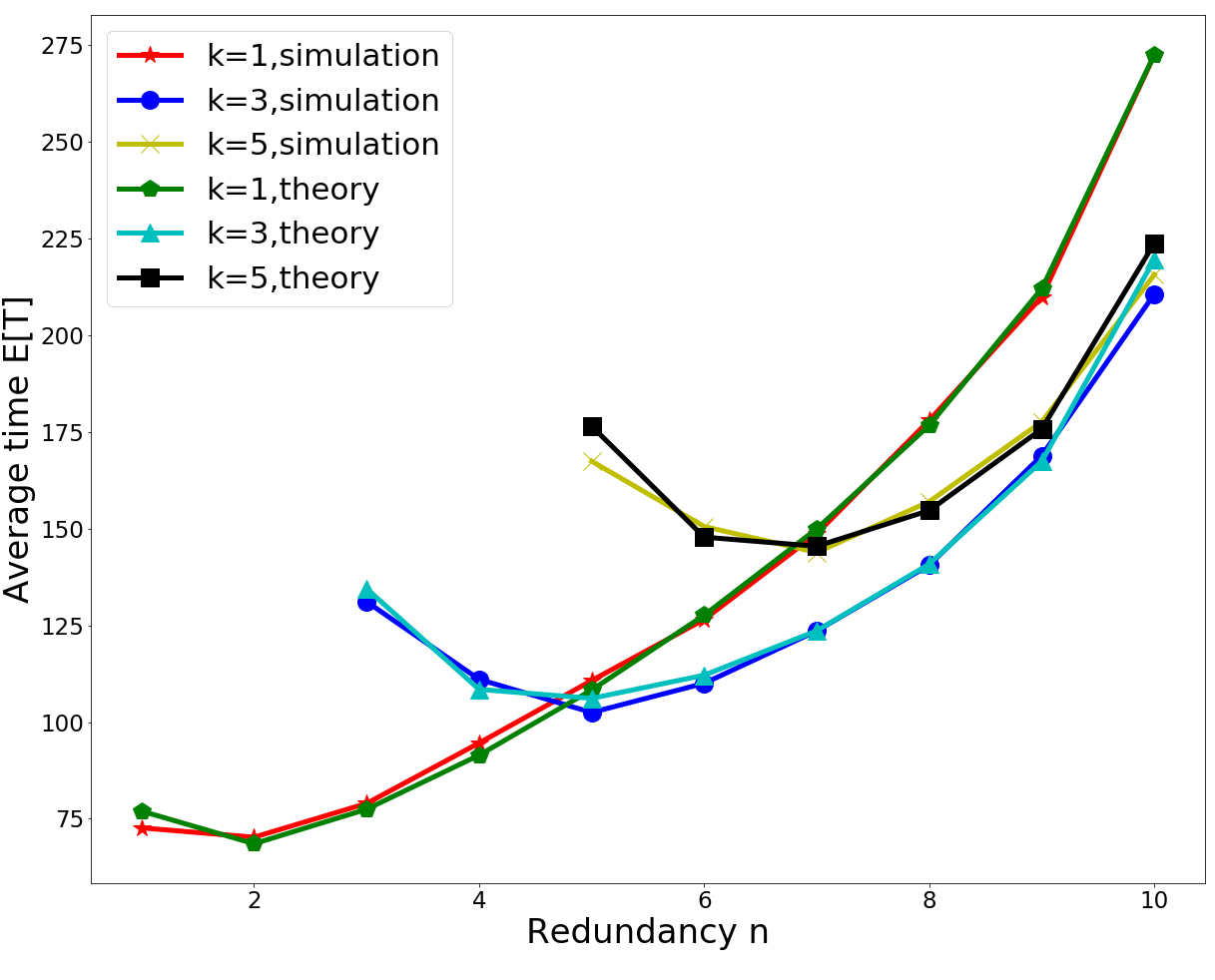}
    \caption{The average time changes with redundancy $n$ under different $k$. The number of relays $r=10$. Both simulation and theoretical results are provided.}
    \label{fig:kredundancy}
\end{figure}

{\it 2) Delay vs.\ $n$ for Different Values of $k$:}
Figure \ref{fig:kredundancy} shows both the simulated and the theoretical results, which closely match each other.
From Fig.~\ref{fig:kredundancy}, we can draw similar conclusions as from Fig.~\ref{fig:mredundancy}. Note that $k=1$, $n=2$ result in the minimum average time. Therefore, splitting the message introduces delay.

\begin{figure}[t]
    \centering
    \includegraphics[scale=0.195]{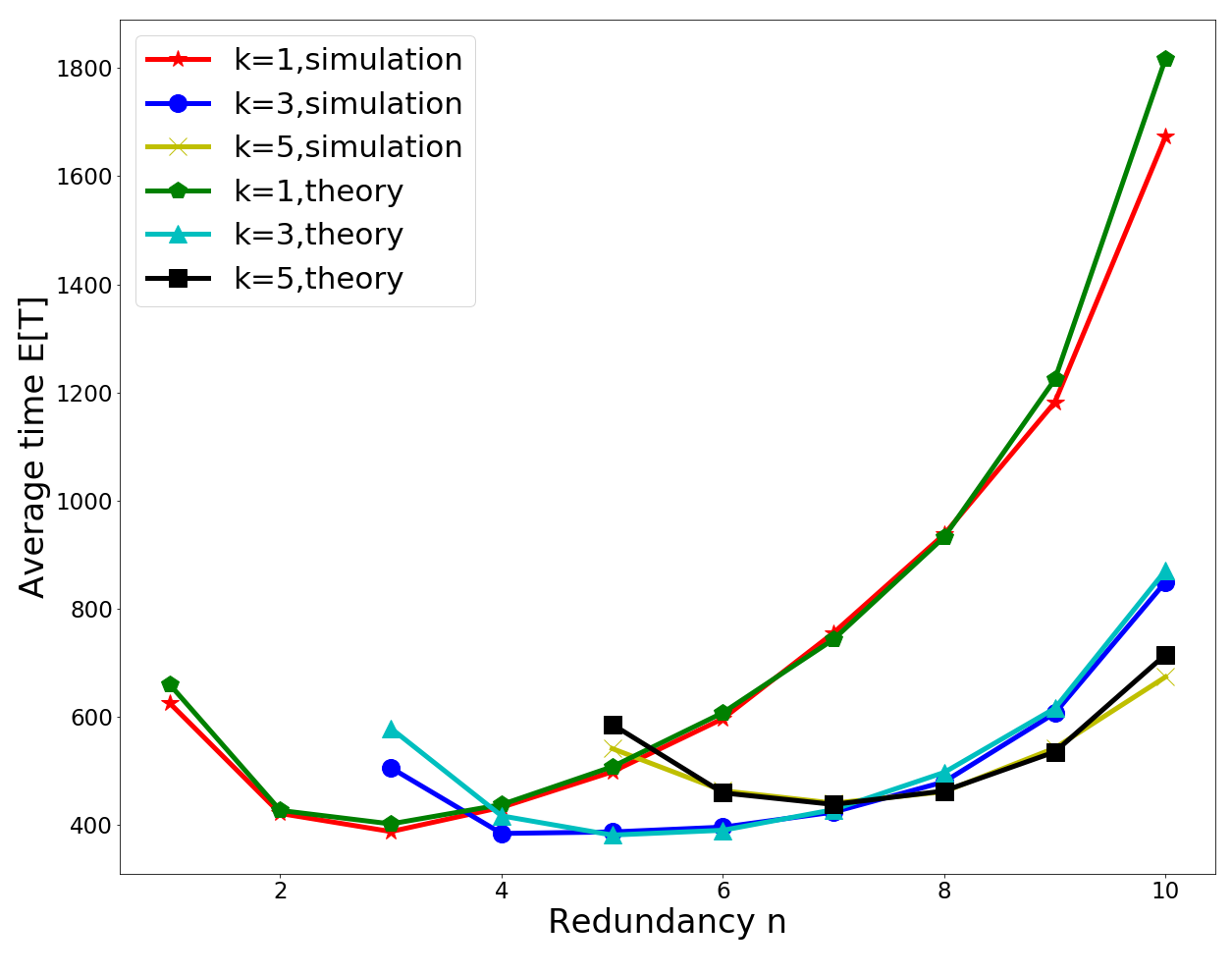}
    \caption{The average time changes with redundancy $n$ under different $k$. The number of relays is $r=10$. Both simulation and theoretical results are provided.}
    \label{fig:kredundancymodel2}
\end{figure}
Figure \ref{fig:kredundancymodel2} shows the results of the second communication delay model.  Comparing with figure \ref{fig:kredundancy}, more redundancy should be introduced to get a lower average time, and when $k=3$ and $n=5$ the average time reaches the minimum. 

Now we can compare the figures from both covertness probability and minimum delay. If we compare figure \ref{fig:covertness} and \ref{fig:kredundancy}, we can find that when $n=1$ and $k=1$, which means the message is not divided and no redundancy is introduced, the covertness communication can get the maximum covertness probability and a small enough average time. However it doesn't mean that the redundancy and chunk transmission are useless. In fact, in the covertness probability simulation, we assume the warden's arrival time follows $U(0,50)$ and the message length is $m=10$, which shows $W=50$ is always larger than $l$. However if the warden arrives more frequently, for example the arrival time follows $U(0,8)$, then $W=8$ is smaller than $l$ when $k=1$. In this case, when $n=1$ and $k=1$, the covertness probability is $0$. To get the overall optimal values of covertness probability and delay, we must introduce some redundancy and divide the message into more data chunks. 

If we simultaneously consider both Fig.~\ref{fig:covertness} and Fig.~\ref{fig:kredundancymodel2}, we see that for $n=1$ and $k=1$ the probability of covertness is high, but so is the expected delay. On the other hand, when $n=5$ and $k=5$, the average time is low, but there is a price to pay in the covertness probability.

\section{Conclusions and Future Work}
We introduced a model for covert message exchange in mobile multi-agent environments. There are four (types of) participants in our communication model: a source (Alice), $r$ relays, a collector (Bob), and a warden (Willie). The model stipulates the participants' mobility patterns and the communications protocols, and defines performance metrics to be the probability of maintaining covertness and the total message transfer time. We obtained expressions for these performance metrics by theoretical derivations and/or simulation, and showed the tradeoff between them as a function of the system parameters.

Many more models for such systems seem reasonable, but have not been studied yet. Future directions could include the exploration of many different ways that the warden can operate but also many different mobility patterns. For example, Willie might need to spend a certain amount of time before being able to detect a transmission at a node, or he could also be adaptive, meaning that he can over time learn which nodes do not have relays and systematically avoid checking them. Regarding, the mobility patterns, random walks on regular graphs or some other area traversal models may be more practical. 

\label{Sec:conclusion}


\section*{Acknowledgments}
This research is based upon work supported by the National Science Foundation 
under Grant No.~SaTC-1816404.

\bibliographystyle{IEEEtran}
\bibliography{biblio.bib}

\begin{thebibliography}{1}
\providecommand{\url}[1]{#1}
\csname url@samestyle\endcsname
\providecommand{\newblock}{\relax}
\providecommand{\bibinfo}[2]{#2}
\providecommand{\BIBentrySTDinterwordspacing}{\spaceskip=0pt\relax}
\providecommand{\BIBentryALTinterwordstretchfactor}{4}
\providecommand{\BIBentryALTinterwordspacing}{\spaceskip=\fontdimen2\font plus
\BIBentryALTinterwordstretchfactor\fontdimen3\font minus
  \fontdimen4\font\relax}
\providecommand{\BIBforeignlanguage}[2]{{%
\expandafter\ifx\csname l@#1\endcsname\relax
\typeout{** WARNING: IEEEtran.bst: No hyphenation pattern has been}%
\typeout{** loaded for the language `#1'. Using the pattern for}%
\typeout{** the default language instead.}%
\else
\language=\csname l@#1\endcsname
\fi
#2}}
\providecommand{\BIBdecl}{\relax}
\BIBdecl

\bibitem{HidingInformationInNoise:BashGT15}
B.~A. Bash, D.~Goeckel, D.~Towsley, and S.~Guha, ``Hiding information in noise:
  Fundamental limits of covert wireless communication,'' \emph{IEEE
  Communications Magazine}, vol.~53, no.~12, pp. 26--31, 2015.

\bibitem{CovertComm:Bloch16}
M.~R. Bloch, ``Covert communication over noisy channels: A resolvability
  perspective,'' \emph{IEEE Transactions on Information Theory}, vol.~62,
  no.~5, pp. 2334--2354, 2016.

\bibitem{DeniableComm:KadheJB14}
S.~Kadhe, S.~Jaggi, M.~Bakshi, and A.~Sprintson, ``Reliable, deniable, and
  hidable communication over multipath networks,'' in \emph{2014 IEEE
  International Symp.\ on Inform.\ Theory}.\hskip 1em plus 0.5em minus
  0.4em\relax IEEE, 2014, pp. 611--615.

\bibitem{girard2004border}
A.~R. Girard, A.~S. Howell, and J.~K. Hedrick, ``Border patrol and surveillance
  missions using multiple unmanned air vehicles,'' in \emph{2004 43rd IEEE
  Conference on Decision and Control (CDC)(IEEE Cat. No. 04CH37601)},
  vol.~1.\hskip 1em plus 0.5em minus 0.4em\relax IEEE, 2004, pp. 620--625.

\bibitem{puri2005survey}
A.~Puri, ``A survey of unmanned aerial vehicles (uav) for traffic
  surveillance,'' \emph{Department of computer science and engineering,
  University of South Florida}, pp. 1--29, 2005.

\bibitem{paley2008cooperative}
D.~A. Paley, F.~Zhang, and N.~E. Leonard, ``Cooperative control for ocean
  sampling: The glider coordinated control system,'' \emph{IEEE Transactions on
  Control Systems Technology}, vol.~16, no.~4, pp. 735--744, 2008.

\bibitem{villa2016development}
T.~Villa, F.~Salimi, K.~Morton, L.~Morawska, and F.~Gonzalez, ``Development and
  validation of a uav based system for air pollution measurements,''
  \emph{Sensors}, vol.~16, no.~12, p. 2202, 2016.

\bibitem{DBLP:conf/infocom/TripathiTM19}
V.~Tripathi, R.~Talak, and E.~Modiano, ``Age optimal information gathering and
  dissemination on graphs,'' in \emph{2019 {IEEE} Conference on Computer
  Communications, {INFOCOM} 2019, Paris, France, April 29 - May 2, 2019}, 2019,
  pp. 2422--2430.

\bibitem{joshi2012coding}
G.~Joshi, Y.~Liu, and E.~Soljanin, ``Coding for fast content download,'' in
  \emph{Communication, Control, and Computing (Allerton), 2012 50th Annual
  Allerton Conference on}.\hskip 1em plus 0.5em minus 0.4em\relax IEEE, 2012,
  pp. 326--333.

\end{thebibliography}
\end{document}